%% file: main.tex
\documentclass{article}

\usepackage{arxiv}

\usepackage{ulem}
\usepackage[utf8]{inputenc} %
\usepackage[T1]{fontenc}    %
\usepackage{hyperref}       %
\usepackage{url}            %
\usepackage{booktabs}       %
\usepackage{amsfonts}       %
\usepackage{nicefrac}       %
\usepackage{microtype}      %
\usepackage{lipsum}		%
\usepackage{epstopdf}
\usepackage{doi}
\usepackage{parskip}
\usepackage{color}
\usepackage{caption}
\usepackage{subcaption}
\usepackage{float}
\usepackage{mathematics}
\usepackage{graphics,graphicx}
\usepackage{wrapfig}
\usepackage{cancel}
\usepackage{mathtools}
\usepackage{mathabx}
\usepackage{enumitem}
\usepackage{todonotes}
\usepackage{soul}
\usepackage{dsfont}
\usepackage{verbatim}
\usepackage{algpseudocode}
\usepackage{algorithm}
\usepackage{tcolorbox}
\usepackage{scalerel}
\usepackage{amssymb}
\usepackage{multirow}

\algnewcommand\And{\textbf{and}}

\newcommand{\MC}[1]{\mathcal{#1}}

\newcommand{\Vtheta}{\Vec{\theta}}
\newcommand{\VTheta}{\Vec{\Theta}}

\newcommand{\rhov}{\widecheck{\rho}}

\newcommand{\WT}[1]{\widetilde{#1}}
\newcommand{\etal}{\textit{et al.}}

\newcommand\includegraphicsifexists[2][width=\linewidth]{\IfFileExists{#2}{\includegraphics[#1]{#2}}{}}

\title{Characterizing Non-Markovian Dynamics of Open Quantum Systems}
\author{ 
\href{https://orcid.org/0000-0001-6882-9737}{\includegraphics[scale=0.06]{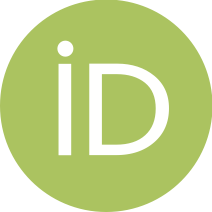}\hspace{1mm}
    Sohail Reddy \thanks{Corresponding author}} \\
	Lawrence Livermore National Laboratory \\
	Livermore, CA 94550 \\
	\texttt{reddy6@llnl.gov} \\
}
\date{}

\begin{document}

\maketitle

\begin{abstract}

    Characterizing non-Markovian quantum dynamics is essential for accurately modeling open quantum systems, particularly in near-term quantum technologies. In this work, we develop a structure-preserving approach to characterizing non-Markovian evolution using the time-convolutionless (TCL) master equation, considering both linear and nonlinear formulations. To parameterize the master equation, we explore two distinct techniques: the Karhunen-Loève (KL) expansion, which provides an optimal basis representation of the dynamics, and neural networks, which offer a data-driven approach to learning system-environment interactions. We demonstrate our methodology using experimental data from a superconducting qubit at the Quantum Device Integration Testbed (QuDIT) at Lawrence Livermore National Laboratory (LLNL). Our results show that while neural networks can capture complex dependencies, the KL expansion yields the most accurate predictions of the qubit’s non-Markovian dynamics, highlighting its effectiveness in structure-preserving quantum system characterization. These findings provide valuable insights into efficient modeling strategies for open quantum systems, with implications for quantum control and error mitigation in near-term quantum processors.
        
\end{abstract}

\input{intro}

\input{formulation}

\input{experiments}

\input{results}

\input{conclusion}

\section*{Acknowledgements}

This work was performed under the auspices of the U.S. Department of Energy by Lawrence Livermore National Laboratory under Contract DE-AC52-07NA27344.

\bibliographystyle{unsrt}
\bibliography{references}

\end{document}

%% file: intro.tex
\section{Introduction} \label{sec:intro}

Performing operations on quantum computers requires precise and accurate control over the quantum device to implement the necessary state transformations reliably. In ideal scenarios, quantum gates should execute deterministic unitary operations, but in real quantum hardware—especially in the Noisy Intermediate-Scale Quantum (NISQ) era—various noise processes, including decoherence, cross-talk, and control errors, introduce deviations from the ideal behavior. These noise processes make it challenging to identify and generate accurate control pulses, as designing any pulse sequence requires precise characterization of the quantum device and modeling of the device dynamics. The complexity of this task increases as quantum systems grow in size because noise is not only present at the level of individual qubits but also manifests in correlated and non-Markovian forms across multi-qubit interactions. Furthermore, imperfections in fabrication lead to device inhomogeneities, requiring calibration protocols tailored to each specific system. Characterizing master equations is crucial for understanding and predicting the behavior of open quantum systems, where interactions with an environment cause decoherence and dissipation. These equations provide a mathematical framework to describe how a system evolves over time, capturing essential features such as memory effects, non-Markovian dynamics, and system-environment correlations. By distinguishing between different types of master equations—such as the Lindblad, Nakajima-Zwanzig, and time-convolutionless forms—researchers can select the most appropriate model for a given physical scenario, ensuring both accuracy and computational efficiency. Without these characterizations, implementing high-fidelity operations becomes infeasible, hindering progress toward large-scale fault-tolerant quantum computation.

Characterization of open quantum systems involves identifying either the process matrix or the generator of the completely positive trace-preserving (CPTP) map. Flynn \etal \cite{Flynn2022} employed an experimental design approach to determine the Hamiltonian of a target system, a methodology later extended by Fioroni \etal \cite{Fioroni2025} and Reddy \etal \cite{Reddy2024:BExD} to calibrate the diagonal Lindblad master equation using experimental data. Rather than focusing on the generator of the CPTP map, White \etal \cite{White2020} and Pollock \etal \cite{Pollock2018} characterized the process tensor to account for non-Markovian dynamics, with the former reporting a device infidelity of $10^{-3}$. Duffus \etal \cite{Duffus2017} demonstrated that improper parameterization of the master equation can degrade device characterization, leading to inaccurate predictions of system dynamics. In such cases, rather than imposing a predefined structure, black-box and grey-box machine learning models have been widely employed for device characterization. Reddy \etal \cite{Reddy2024:QUDE} introduced a technique that enhances known dynamics (modeled by a master equation) with a machine learning agent to capture latent dynamics. Youssry \etal \cite{Youssry2020} applied a grey-box approach, integrating neural networks with quantum mechanical models to characterize noise and control open quantum systems. Goswami \etal \cite{Goswami2021} used supervised learning to identify non-Markovianity with 90\% accuracy, even with tomographically incomplete measurements. Genois \etal \cite{Genois2021} leveraged a structure-preserving recurrent neural network to improve characterization, calibration, and control. Krastanov \etal \cite{Krastanov2020} employed an interpretable, structure-preserving machine learning model to characterize the non-Markovian dynamics of a qubit coupled to a readout cavity.

This work considers a `grey-box' approach where the completely positive trace-preserving (CPTP) property is maintained through the structure of the master equation while allowing for arbitrary parameterization of the process governing unitary and decoherence dynamics. To enable modeling and characterization of both Markovian and non-Markovian dynamics, we employ the time-convolutionless (TCL) master equation. This approach ensures structure preservation and interpretability of the learned model, allowing it to distinguish between unitary and dissipative dynamics. The TCL master equation is parameterized by a basis of linear operators, with process coefficients that are learned from device data. The process coefficients are modeled using the Karhunen-Loève (KL) decomposition, which provides a low-rank approximation of the process tensor, as well as Neural Network models. Furthermore, we consider both the linear TCL equation, where the coefficients are independent of the quantum state, and the nonlinear TCL equation, where the coefficients depend on the quantum state. We apply this approach to identify latent dynamics in a superconducting quantum processing unit (QPU) at the Quantum Design and Integration Testbed (QuDIT) at Lawrence Livermore National Laboratory (LLNL).

%% file: formulation.tex
\section{Non-Markovian Master Equation} \label{sec:mastereq}

The Nakajima-Zwanzig (NZ) equation is a fundamental integro-differential equation used in open quantum systems to describe the reduced dynamics of a system interacting with an environment. It is derived using the projection operator technique, resulting in a memory kernel formulation that accounts for non-Markovian effects:
\eq{NZ}{
\ddt{\rho} = \integral{0}{t}{\mathcal{K}(t,t') \rho(t')~dt'} ,
}
where $\mathcal{K}(t,t')$ is the memory kernel encoding system-environment correlations. Unlike time-local master equations, the NZ equation retains past influences, making it more accurate for strong coupling and highly non-Markovian dynamics, though often computationally inefficient. Nestmann \etal \cite{Nestmann2021} showed how the memory effects in non-Markovian dynamics are manifested in a time-local way. 

The time-convolutionless (TCL) master equation is a formalism used in open quantum systems to describe the reduced dynamics of a system interacting with an environment. Unlike the Nakajima-Zwanzig equation, which involves a memory kernel and an integro-differential form, the TCL master equation expresses the evolution as a time-local differential equation, making it more convenient for practical computations. The time-convolutionless (TCL) master equation is defined as
\eq{TCL}{
    \ddt{\rho} = -i \sum_{i=1}^{N^2 - 1}  \omega_i(t) \sbrac{\Lambda_i,\rho} + \sum_{i=1}^{N^2 - 1} \gamma_i(t) \rbrac{\Lambda_i \rho \Lambda^\dagger_i - \dfrac{1}{2}\cbrac{\Lambda^\dagger_i \Lambda_i, \rho}}
}
where the first and second terms represent the unitary and disspative terms, respectively. The master equation can be vectorized into a Liouvillian superoperator form $\ddt{\rhov}=\MC{L}(t) \rhov$, with $\rhov := \mathrm{vec}(\rho)$ and
\eq{TCL:Liouvillian}{
    \MC{L}(t) = -i \sum_{i=1}^{N^2 - 1} \omega_i(t)\rbrac{\Lambda_i \otimes I - I \otimes  \Lambda^T_i} + \gamma_i(t) \rbrac{ \Lambda_i \otimes \Lambda^*_i  - \dfrac{1}{2} \rbrac{ \Lambda^\dagger_i \Lambda_i \otimes I + I \otimes  (\Lambda^\dagger_i \Lambda_i)^T } }
}
which has the solution 
\eq{TCL:Solution}{
     \rhov(t) = \underbrace{\MC{T} \exp\rbrac{\integral{t_0}{t}{\MC{L}(\tau)~d\tau}}}_{\mathbb{L}(t_0,t)} \rhov(t_0)
}
where $\MC{T}$ is the time-ordering operator.
Since inaccurate parameterization may lead to inaccurate characterization, we consider the following general form of the TCL
\eq{TCL:Liouvillian:Gen}{
    \ddt{\rho} = -i \sum_{i=1}^{N^2 - 1}  \omega_i(t) \sbrac{\Lambda_i,\rho} + \sum_{i,j=1}^{N^2 - 1}  \Gamma_{ij}(t) \rbrac{\Lambda_i \rho \Lambda^\dagger_j - \dfrac{1}{2}\cbrac{\Lambda^\dagger_i \Lambda_j, \rho}}
}
where $\Gamma$ is a positive semi-definite matrix of decoherence rates defined by its Cholesky decomposition, $\Gamma = Q Q^T$, where $Q$ is an upper triangular matrix. 
The linear operator $\mathbb{L}(t_0,t): \rhov(t_0) \mapsto \rhov(t)$ is a completely-positive-trace-preserving (CPTP) map and satisfies the quantum dynamical semigroup properties:
\begin{enumerate}
    \item $\mathbb{L}(t_0,t_0) \rhov(t_0) = \rhov(t_0)$
    \item $\mathbb{L}(t_0,t+\tau) = \mathbb{L}(t,\tau)\mathbb{L}(t_0,t)$
    \item $\mathbb{L}(t_0,\tau)$ is completely positive for all $\tau \geq t_0$
\end{enumerate}

\section{Characterizing Master Equation} \label{sec:characterization}

Quantum characterization amounts to identifying parameters in a mathematical ansatz (e.g., a master equation) that best fit experimental measurement. Experimental data can be obtained via interferometry protocols - such as Rabi, Ramsey and Hahn-Echo - or from optimal experimental design \cite{Reddy2024:BExD}. 

Consider experimental data $\cbrac{\widetilde{\rho}(t_i)}_{i=1}^{N_T}$. Here we consider $\R^{2(N^2-1)}\ni \Vtheta = \bigcup_{i=1}^{N^2 - 1} \cbrac{\omega_i, \gamma_i}$
Then, the characterization problem amounts to solving the following constrained minimization problem:
\aligneq{TCL:ConstOptProb}{
    \Vtheta^\star := \arg \min_{\Vtheta \in \VTheta}& \underbrace{ \sum_{i=1}^{N_T}~\Norm{ \widetilde{\rho}(t_i) - \rhov(t_i;\Vtheta) }^2}_{\MC{C}(\Vtheta)}, \\
                \mathrm{subject~to}: & \ddt{\widecheck{\rho}} =  \MC{L}(t;\Vtheta) \rhov. \\
}
The constraint can be incorporated directly into the cost function leading to the following unconstrained optimization problem
\aligneq{TCL:OptProb}{
    \Vtheta^\star &:= \arg \min_{\Vtheta \in \VTheta} \underbrace{ \sum_{i=1}^{N_T}~\Norm{ \widetilde{\rho}(t_i) - \mathbb{L}(t_0,t_i;\Vtheta)\rhov(t_0)}^2}_{\MC{C}(\Vtheta)} \\
}
The cost function in \eqref{TCL:OptProb} requires time-evolution of the state over the entire training interval, which is performed sequentially.

\subsection{Karhunen-Loeve Expansion Parameterization} \label{subsec:KLE}

The Karhunen-Loève (KL) expansion is a mathematical technique used to represent a stochastic process as an infinite series of orthogonal functions weighted by uncorrelated random variables. It provides an optimal basis for decomposing a random field, similar to how the Fourier series represents deterministic functions. Given a stochastic process $\eta(t)$ with a well-defined covariance function, $k(t,t')$, the KL expansion expresses it as
\eq{KL}{
\eta(t) = \overline{\eta} + \sum_{i=0}^\infty \eta_i \phi_i(t)
}
where $\overline{\eta}$ is the process mean, $\phi_i$ are the eigenfunctions of the covariance kernel, $k(t,t')$, and $\eta_i$ are uncorrelated random coefficients with mean zero and variances given by the corresponding eigenvalues. This expansion allows for efficient dimensionality reduction by truncating the series while preserving most of the variance in the process. Its optimality in terms of mean-square error makes it particularly valuable for representing complex random fields in computational simulations. Since analytical expressions for the eigenfunction are only known for a few select kernels, discrete approximations are often employed in practice. We consider each element of $\Vtheta$, that is $\omega_i$ and $\gamma_i$, to be a stochastic process described by their KL expansions. Then, characterization amounts to identifying optimal expansion coefficients. 

Here, we only consider kernels for which analytical expressions are known, namely, the exponential and squared-exponential kernels. The exponential kernel is defined as \cite{Maitre2010}
\[
k(t,t') = \sigma^2 \exp\rbrac{-\dfrac{|t-t'|}{2 \kappa^2}}
\]
and has eigenfunctions, $\phi_i$ defined by
\[
\phi_i(t) = 
\begin{cases}
\dfrac{\cos[\omega_i(x-1/2)]}{\sqrt{\dfrac{1}{2} + \dfrac{\sin(\omega_i)}{2\omega_i}}},& \text{if $i$ is even}\\
\\
\dfrac{\sin[\omega_i(x-1/2)]}{\sqrt{\dfrac{1}{2} - \dfrac{\sin(\omega_i)}{2\omega_i}}}, & \text{if $i$ is odd}
\end{cases}
\]
and eigenvalues, $\lambda_i$, defined as
\[
\lambda_i = \sigma^2 \dfrac{\kappa^2}{1 + (\omega_i \kappa)^2}
\]
where $\omega_i$ are the (ordered) positive roots of the characteristic equation
\[
(1-\kappa \omega \tan(\omega/2))(\kappa \omega  + \tan(\omega/2)) = 0
\]

The squared exponential kernel is defined as \cite{Rasmussen2006}
\[
k(t,t') = \sigma^2 \exp\rbrac{-\dfrac{(t-t')^2}{2 \kappa^2}}
\]
and has eigenvalues, $\lambda_i$, and eigenfunctions, $\phi_i$, defined by
\[
\lambda_i = \sqrt{\dfrac{2a}{A}}B^k, \quad ~\phi_i(t) = \exp(-(c-a)x^2) H_i(\sqrt{2c}x)
\]
where $H_i(x)$ is the $i$th order Hermite polynomial, $a^{-1}=4\sigma^2$, $b^{-1}=2 \kappa^2$ and 
\[
c=\sqrt{a^2 + 2ab}, \quad ~A=a+b+c, \quad~B=b/A.
\]
In practice, we truncate the expansion at $M$ terms. Furthermore, we denote the $i$th expansion coefficient for $\theta \in \Vtheta$ as $\eta_i^{(\theta)}$. For simplicity, we truncation the expansion for each $\theta \in \Vtheta$ at $M$ terms. We refer to the KL parameterization using the Exponential and Squared-Exponential covariance kernel as \emph{KL-Exp.} and \emph{KL-Sq. Exp.}, respectively.

\subsection{Neural Network Parameterization} \label{subsec:DNN}

Although the TCL master equation in its time-local form models non-Markovian dynamics (with appropriate parameterization of the process coefficient), the model is linear. Structure-preserving nonlinearity can be introduced through appropriate parameterization of the process coefficients. To introduce nonlinearity, we consider a nonlinear function $\MC{N}:(\rhov,t) \mapsto \Vtheta$, modeled as a feed-forward neural network. Then, the mapping $\mathcal{N}(\rho)$ is an $L$-fold composition of network layers $\mathcal{N}_\ell$
\eq{DNN}{
	\mathcal{N}(\rhov,t) := \mathcal{N}_{L} \circ \mathcal{N}_{L-1} \circ \ldots \circ \mathcal{N}_{1} (\rhov,t)
}
where each layer $\mathcal{N}_\ell: \R^{N_\ell} \rightarrow \R^{N_{\ell+1}}$ consists of an affine transformation followed by a nonlinear activation $\sigma : \R \rightarrow \R$ that is applied component-wise:
\eq{NetworkLayer}{
	\mathcal{N}_\ell(\mathbf{x}) = \sigma_l \left( \mathbf{W}_\ell \mathbf{x} + \mathbf{b}_\ell  \right)
}
where $\mathbf{W}_\ell$ and $\mathbf{b}_\ell$ are the weight matrix and bias vector, respectively, for layer $\ell$. More precisely, the density matrix is represented by expansion coefficients in some basis. Hence, for the dimensionality, $d$, of the Hilbert space of the quantum state, we have $\MC{N}:\R^{d^2} \rightarrow \R^{2(d^2-1)}$. The trainable parameters of this ansatz are the weights, $\mathbf{W}_\ell$, and biases, $\mathbf{b}_\ell$, of each layer of the neural network. In the absence of the nonlinear activation function (i.e. $\sigma$ is the identity), then the $L$-fold composition reduces to an affine map. Consider the case of an affine map $\Vtheta = \mathbf{W}\mathbf{x} + \mathbf{b}$, with the weight block matrix 
\[
    \mathbf{W} = \rmatrix{ \mathbf{A} & b \\ \mathbf{C} & d}, ~ \mathbf{x} = \smatrix{\rhov \\ t}, ~\mathrm{with}~\mathbf{A},\mathbf{C} \in \R^{N \times N},~\mathrm{and}~b,d \in \R.
\]
Then, under certain conditions, we have the following generalization of the TCL equations:
\begin{enumerate}
    \item If $\rhov \in \mathrm{null}(\mathbf{A})$ and $\rhov \in \mathrm{null}(\mathbf{C})$, then $\MC{N}(\rhov,t)=\MC{N}(t)$ and the model reduces to the linear TCL master equation
    \item If $\mathbf{x}  \in \mathrm{null}(\mathbf{W})$, then $\MC{N}(\rhov,t)= \mathbf{b}$ and the model reduces to the time-independent, Markovian Lindblad master equation
\end{enumerate}
We investigate both nonlinear model as well as an affine model both using the tanh activation function. We refer to the parameterization by an affine map and neural network map as \emph{Affine} and \emph{Neural Network}, and refer to the master equation for which the density matrix is an input to the parameterization as the nonlinear TCL master equation.

%% file: experiments.tex
 \section{Characterizing Dynamics of LLNL Testbed's QPUs} \label{sec:qudit}

We apply our technique to characterize the dynamics of a quantum processing units (QPU) at LLNL's Quantum Design and Integration Testbed (QuDIT). The qubit is a 2D single Tantalum transmon on a sapphire substrate \cite{Place2021} mounted at 10 mK in a dilution fridge.
The parameters, such as transition frequencies ($\omega_{01}$), energy decay times ($T_1$), and dephasing times ($T_2$) of the qubit were estimated using standard characterization protocols, using standard Rabi, Ramsey, and energy decay measurements. The Hamiltonian for superconducting transmon, in the rotating frame is given as
\eq{Hamiltonian}{
  H(t) := \underbrace{ \left( \omega - \omega^{rot} \right) a^\dagger a }_{H_s} + \underbrace{ p(t) \left(a + a^\dagger \right) + iq(t) \left(a - a^\dagger\right) }_{H_c}
}
where $H_s$ and $H_c$ are the system and control Hamiltonians, respectively, $\omega^{rot}$ is the frequency of rotation, with parameters
\[
\omega = 3.448~\mathrm{GHz},~ T_1 = 214~\mu s,~\mathrm{and} ~T_2 =32 ~\mu s
\]
Here, the dynamics of the system are described by the following Lindblad master equation
\eq{QuditEq}{
    \ddt{\rho} := -i[H(t),\rho] + \sum_{i=1}^{2} \dfrac{1}{T_i} \rbrac{L_i \rho L_i^\dagger - \dfrac{1}{2} \left\lbrace L_i^\dagger L_i , \rho \right\rbrace}
}
with 
\expression{
L_1 = \rmatrix{0 & 1 \\ 0 & 0}, \mathrm{~and}\quad L_2 = \rmatrix{0 & 0 \\ 0 & 1}
}

The training and validation data sets are generated by applying constant square pulses in the rotating frame at $\omega^{rot} = \omega_{01}$. The set of experiments $\Vec{\xi}$ consists of various pulse amplitudes $\xi_j \sim \mathcal{U}(0,p_{max})$ with a maximum amplitude of $p_{max} = 3.47$~MHz with $q(t)=0$. The pulses are applied for a total duration of $T=50~\mu$s with measurements sampled every $4~$ns. A total of 5000 shots are performed and the measured states are classified using a Gaussian Mixture Model (GMM) from which, the density matrices at each timestep are estimated by quantum state tomography using linear inversion estimate (LIE)~\cite{Qi:2013}. The density matrices estimated using LIE may not be valid quantum states. To obtain a valid density matrix (i.e., one that is positive semi-definite with $\mathrm{Tr}(\rho)=1$) we apply a spectral filter with renormalization. In particular, we construct filtered density matrices as
\begin{align} \label{eq:SpectralFilter}
	\WT{\rho} = \sum_{i=1}^N \mathcal{E}_i | \Psi_i \rangle \langle \Psi_i | 
    \quad \text{with} \quad
	\mathcal{E}_i = \dfrac{\mathcal{H}(E_i)\cdot E_i}{\sum_j^N \mathcal{H}(E_j)\cdot E_j},
\end{align}
where $\cbrac{\Psi_i}_{i=1}^N$ and $\cbrac{\mathcal{E}_i}_{i=1}^N$ are the eigenvectors and the corresponding filtered and renormalized eigenvalues, respectively, $\cbrac{E_i}_{i=1}^N$ are the original (unfiltered) eigenvalues and the filter function $\mathcal{H}(x)$ is the Heaviside step-function.

%% file: results.tex
\subsection{Problem Formulation} \label{subsec:problemform}

Denote by $\boldsymbol{\xi}$, a set of experiments that are parameterized by control functions which determine the time-evolution of the true/exact quantum state. For the examples below, this set consists of various microwave control pulses, $p(t), q(t)$ in \eqref{Hamiltonian} which will be described in Section \ref{sec:qudit}. Each experiment defines thetrue/exact evolution of the density matrix, $\widetilde{\rho}(t_j; \xi_i)$, which is estimated using quantum state tomography \cite{Qi:2013} at various time steps $t_j \in (0,T]$.

Let $\boldsymbol{\vartheta}$ denote the set of learnable parameters in the mathematical ansatz, that is,
\[
\boldsymbol{\vartheta} = \left\lbrace \eta_i^{(\alpha_l)} \bigcup \eta_i^{(\gamma_l)}: l=1,\ldots,N^2-1,~i=0,\ldots,M \right\rbrace
\] 
for the KL parameterization in Section \ref{subsec:KLE} and $\boldsymbol{\vartheta} = \left\lbrace \mathbf{W}_l \bigcup \mathbf{b}_l : l = 1,\ldots, L \right\rbrace$ for the neural network model defined in Section \ref{subsec:DNN}. Furthermore, consider a KL mapping $\mathcal{F}_{KL}: \boldsymbol{\vartheta} \mapsto \boldsymbol{\theta}$ given by \eqref{KL} and a neural network mapping $\mathcal{F}_\MC{N}: \boldsymbol{\vartheta} \mapsto \boldsymbol{\theta}$ given by \eqref{DNN}.
Then, the training procedure finds optimal parameters $\boldsymbol{\vartheta}$ such that the resulting augmented differential equation matches the measured data $\widetilde{\rho}$ at time-steps $t_j$. To this end, the following constraint optimization problem is solved: 
\subeqs{OptimizationProb}{
	\begin{align}
		\begin{split} \label{eq:LossFunction}
			\boldsymbol{\vartheta}^\star := & \arg\min_{\boldsymbol{\vartheta} \in \boldsymbol{\Theta}} ~~ \sum_{i=1}^{|\boldsymbol{\xi}|}  \mathcal{C}(\MC{F}(\boldsymbol{\vartheta});\xi_i)
		\end{split}\\
		\begin{split} \label{Eq:Constraint}
			\mathrm{subject~to~} &: \dfrac{\partial \rho}{\partial t} = -i \left[H_c(\xi_i), \rho \right] + \mathcal{L}(t;\MC{F}(\boldsymbol{\vartheta}))\rho, \ \forall t \in (0,T] 
		\end{split}
	\end{align}	
}
where $\MC{F}(\boldsymbol{\vartheta}) = \MC{F}_{KL}(\boldsymbol{\vartheta})$ for the KL parameterization, $\MC{F}(\boldsymbol{\vartheta}) = \mathcal{F}_\MC{N}(\boldsymbol{\vartheta})$ for the neural-network parameterization.
The control Hamiltonian $H_c(\xi_i)$ in \eqref{Constraint} encode the experimental setup. 
and take $\Lambda_i$ to be the upper triangular part of the Gellmann matrices.

In order to evaluate the objective function, the dynamical system \eqref{TCL} is evolved numerically, using an explicit 4th-order Runge-Kutta scheme, for the given experimental setup in $\boldsymbol{\xi}$ and the current parameters $\Vec{\theta}$. This numerical evolution yields the predicted evolution $\cbrac{\rho(t_j)}_{j=1}^{N_T}$ which is compared to the true evolution defined by various forms of the loss function. 
We utilize the Julia SciML \cite{rackauckas2017differentialequations} framework to solve this optimization problem with the ADAM as an initial optimizer, followed by L-BFGS optimizer. The gradient is calculated using automatic differentiation. 

After training on a defined set of experiments and time-horizon $0<t_j<T$, we investigate the efficacy of the trained augmented model on a set of validation experiments. In particular, we show that the trained model is able to accurately predict the time-evolution on time domains much longer than the training domain, for experimental setups that have not been included in training. 
We refer to the accuracy over the training and validation sets as \emph{interpolation} and \emph{extrapolation} accuracy, respectively.

In the sections that follow, we will compare the solutions obtained using different formulations using the trace distance $T(\widetilde{\rho},\rho) = \frac{1}{2}\mathrm{Tr}\left[ \sqrt{(\widetilde{\rho}-\rho)^\dagger (\widetilde{\rho}-\rho)} \right]$. %
In each case, we will apply the spectral filter in Eq. \ref{eq:SpectralFilter} to obtain a valid density matrix.

\subsection{Characterizing Noisy QPUs} \label{subsec:results}

\begin{figure}[h] \centering
\begin{subfigure}[h]{\textwidth} %
\includegraphicsifexists[width=0.33\textwidth]{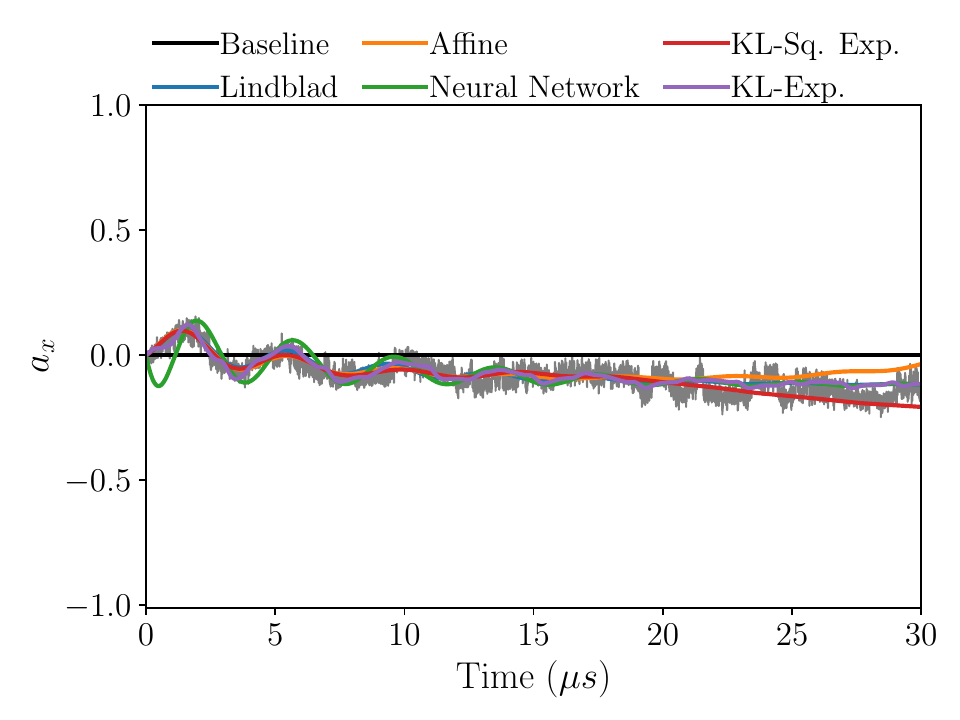} 
\includegraphicsifexists[width=0.33\textwidth]{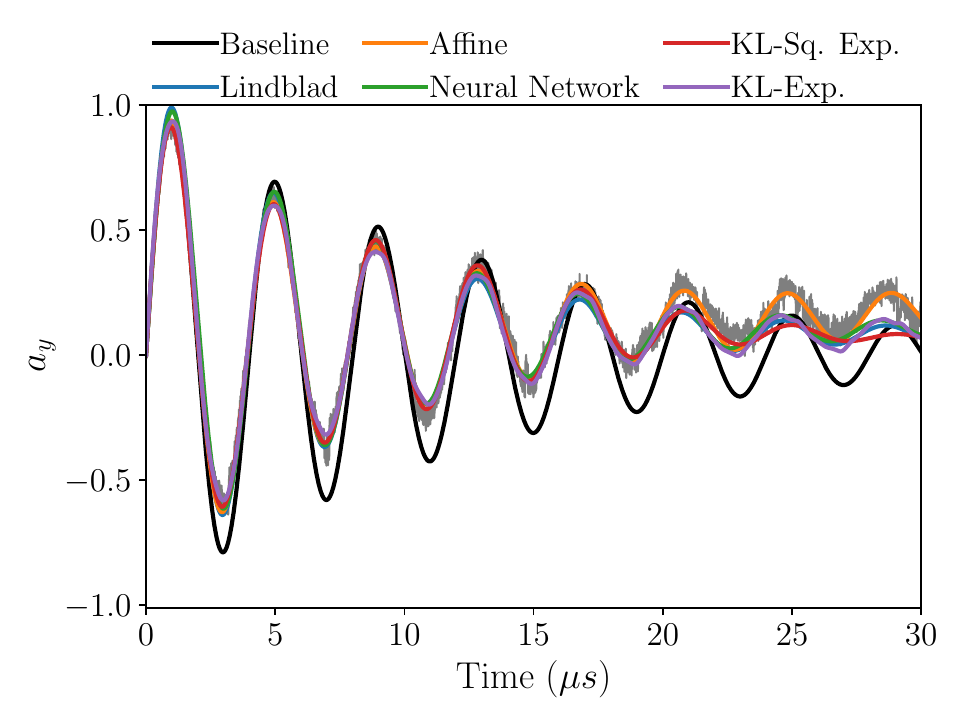}
\includegraphicsifexists[width=0.33\textwidth]{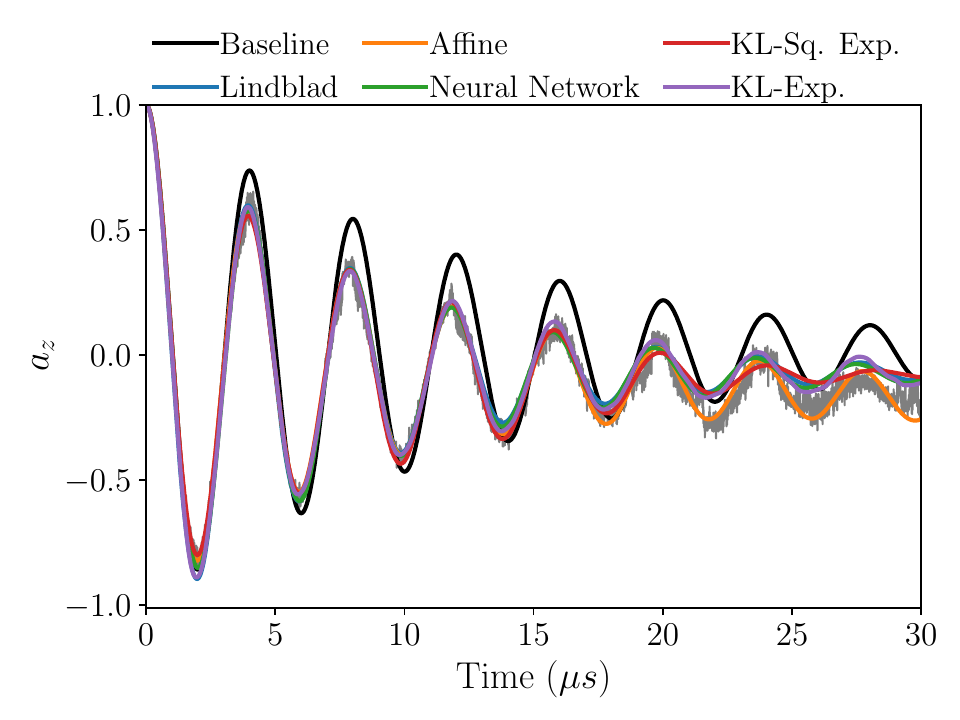}
\caption{}
\label{fig:TCL:bloch0}
\end{subfigure}
\begin{subfigure}[h]{\textwidth} %
\includegraphicsifexists[width=0.33\textwidth]{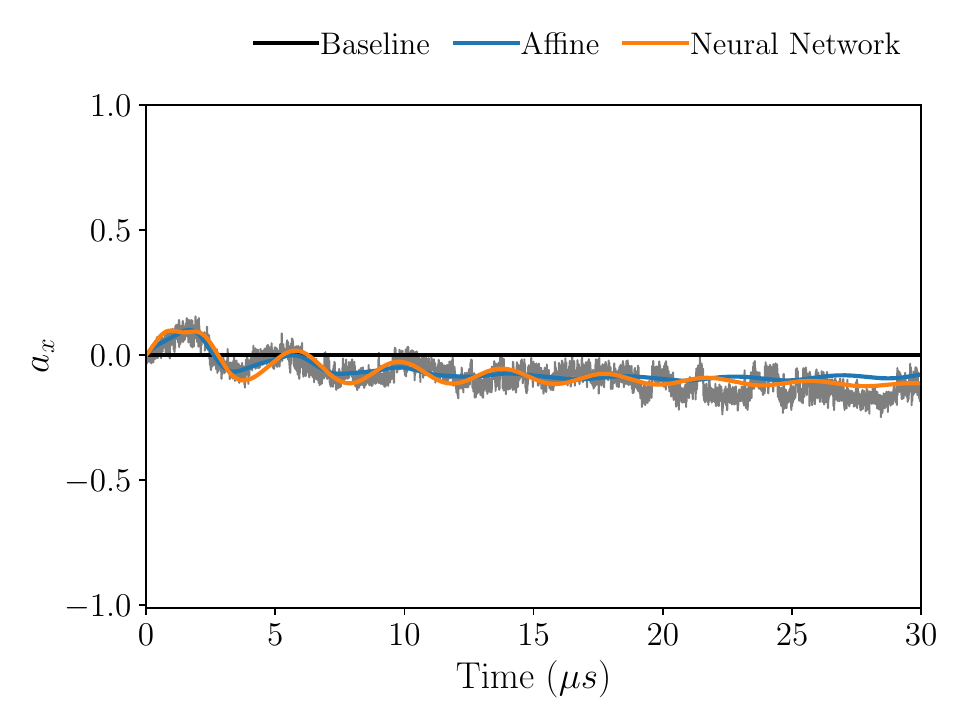} 
\includegraphicsifexists[width=0.33\textwidth]{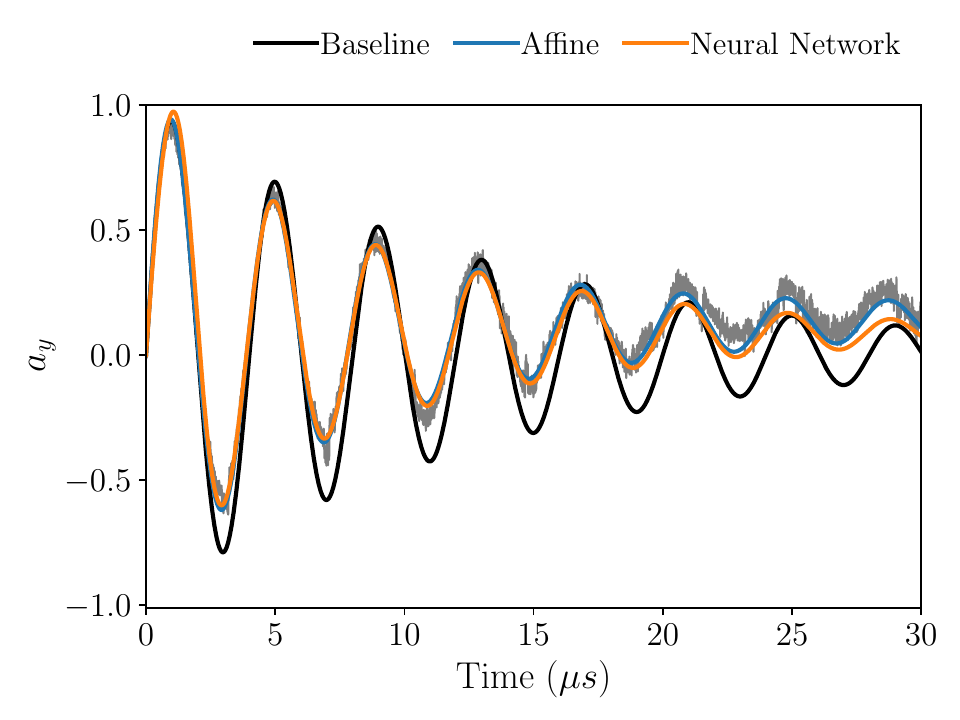} 
\includegraphicsifexists[width=0.33\textwidth]{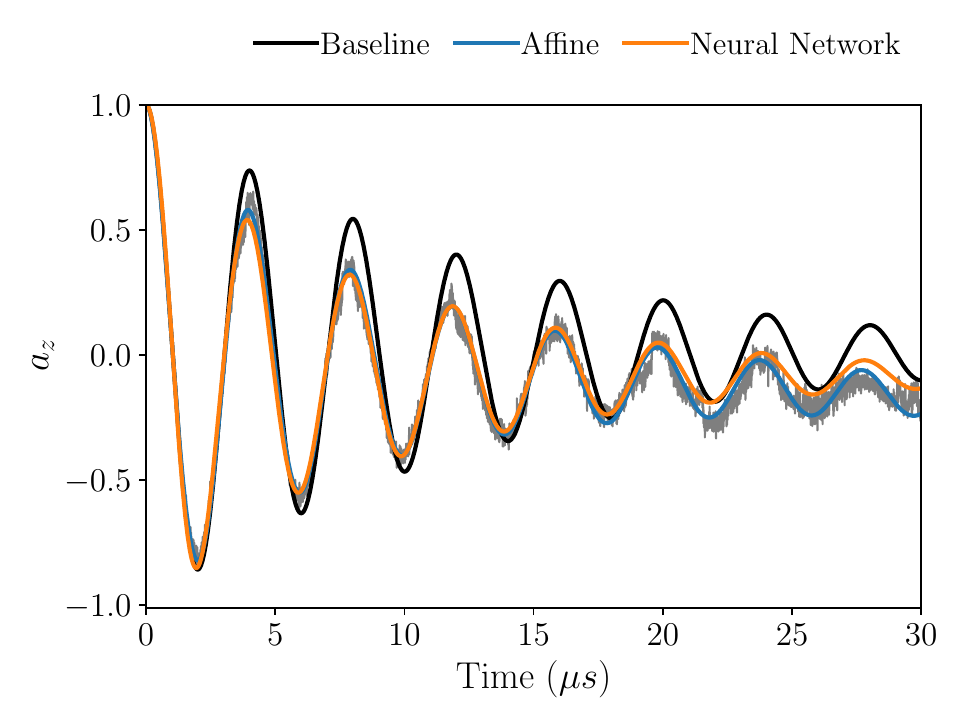} 
\caption{}
\label{fig:NTCL:bloch0}
\end{subfigure}
\captionsetup{singlelinecheck=off,font=footnotesize}
\caption[]{Evolution of the bloch vector components corresponding to the density matrices predicted by the (top) linear TCL and (bottom) nonlinear TCL master equations using various parameterization of the process coefficients.}
\label{fig:TCL:BlochComponents}
\end{figure}

We apply our characterization technique to identify the non-Markovian unitary and decoherence dynamics of a QPU on QuDIT at LLNL. We compare our technique to two benchmarks: 
\begin{enumerate}
    \item \emph{Baseline}: We compute the transition frequency and the $T_1$ and $T_2$ decoherence times using Ramsey interferometry and energy decay experiments. Here, the underlying model is time-independent, with the Lindblad master equation given by \eqref{QuditEq}. 
    \item \emph{Lindblad}: Here we consider the model parameters, $\Vtheta$, in the optimization problem \eqref{OptimizationProb} to be time-independent, which describes the generalized, Markovian dynamics.
\end{enumerate}
We compare the various models using the bloch vector representation of the computed density matrices. The density matrices of a qubit can be represented in the Pauli bases, $\sigma = \cbrac{\sigma_x,\sigma_y,\sigma_z}$,
\expression{
\rho = \frac{1}{2} I + \sum_{i=1}^3 a_i \sigma_i
}
where $a_i = \mathrm{Tr}(\rho \sigma_i)$ is the bloch vector component corresponding to the Pauli matrix, $\sigma_i$, and $I$ is the identity matrix. Figure \ref{fig:TCL:BlochComponents} shows the bloch vector components obtained using the various parameterization of the process coefficients and master equations. We see that for the linear TCL, the Affine parameterization better predicts the evolution of the $a_y$ and $a_z$ components, but is less accurate for $a_x$, particularly when extrapolating outside the training window. Both the Affine and KL-Sq. Exp. parameterization are similar within the training interval, with the Affine parameterization outperforming all others in extrapolation. Note that the affine parameterization is a subset of the KL-Sq. Exp. parameterization as the first two Hermite basis exactly yield an affine map. Hence, the deterioration in extrapolation accuracy is primarily attributed to the higher-order terms in the KL expansion. Nevertheless, all methods outperform the standard characterization by Ramsey interferometry. The Affine and neural network models for the nonlinear TCL master equation exhibit similar accuracy, in both interpolation and extrapolation, to those for the linear TCL equation. 
 
\begin{figure}[h] \centering
\begin{subfigure}[h]{0.33\textwidth} %
\includegraphicsifexists[width=\textwidth]{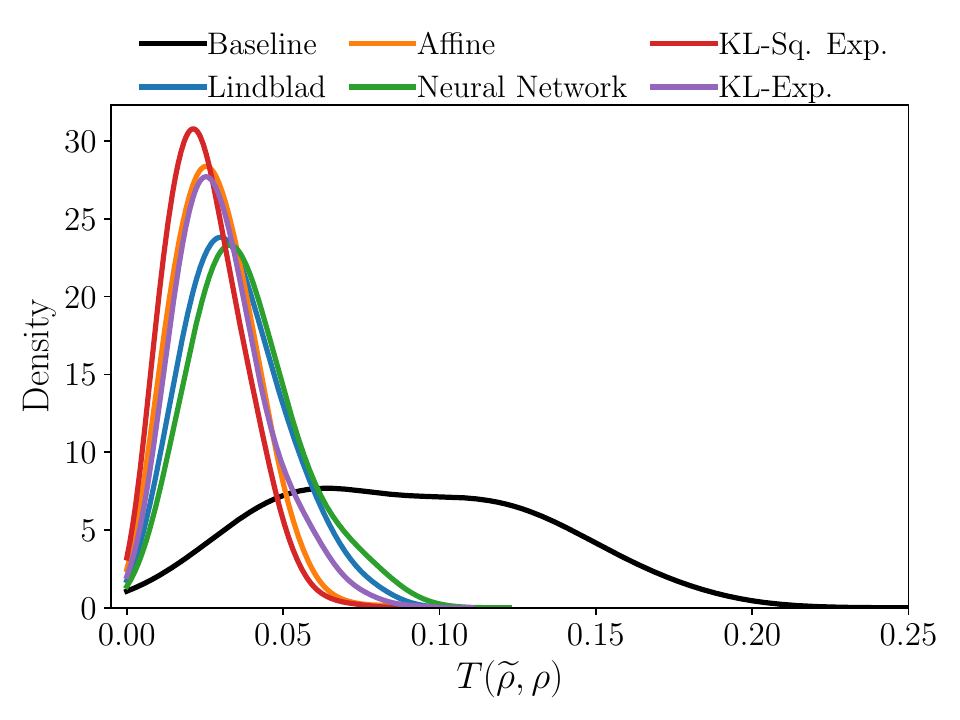} \caption{}\label{fig:TCL:TrDist:Int}
\end{subfigure}
\begin{subfigure}[h]{0.33\textwidth} %
\includegraphicsifexists[width=\textwidth]{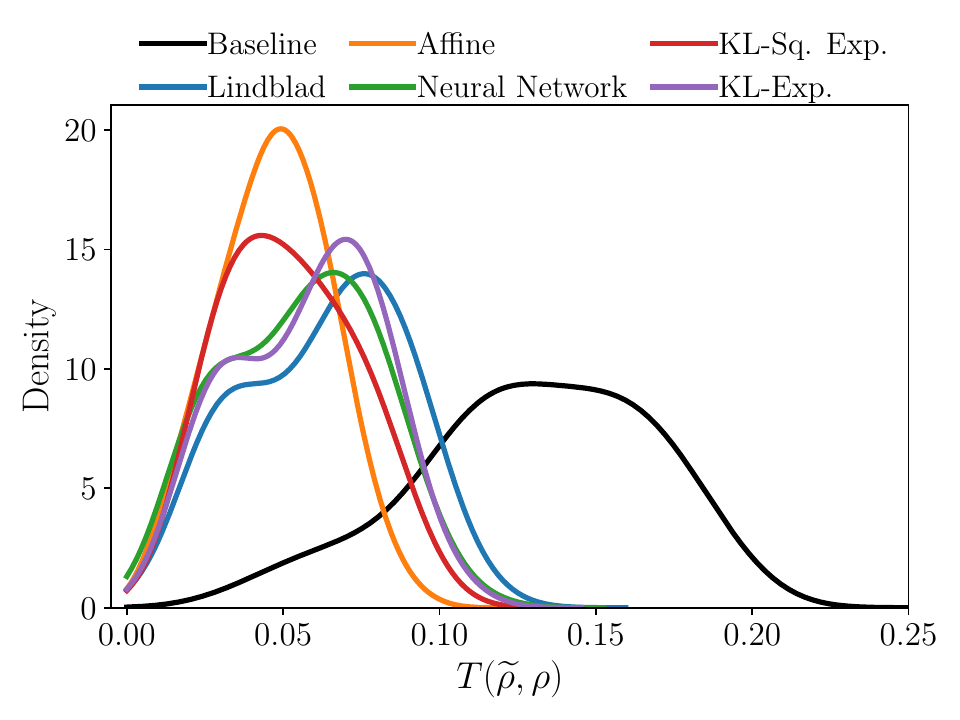} \caption{}\label{fig:TCL:TrDist:Ext}
\end{subfigure}
\begin{subfigure}[h]{0.33\textwidth} %
\includegraphicsifexists[width=\textwidth]{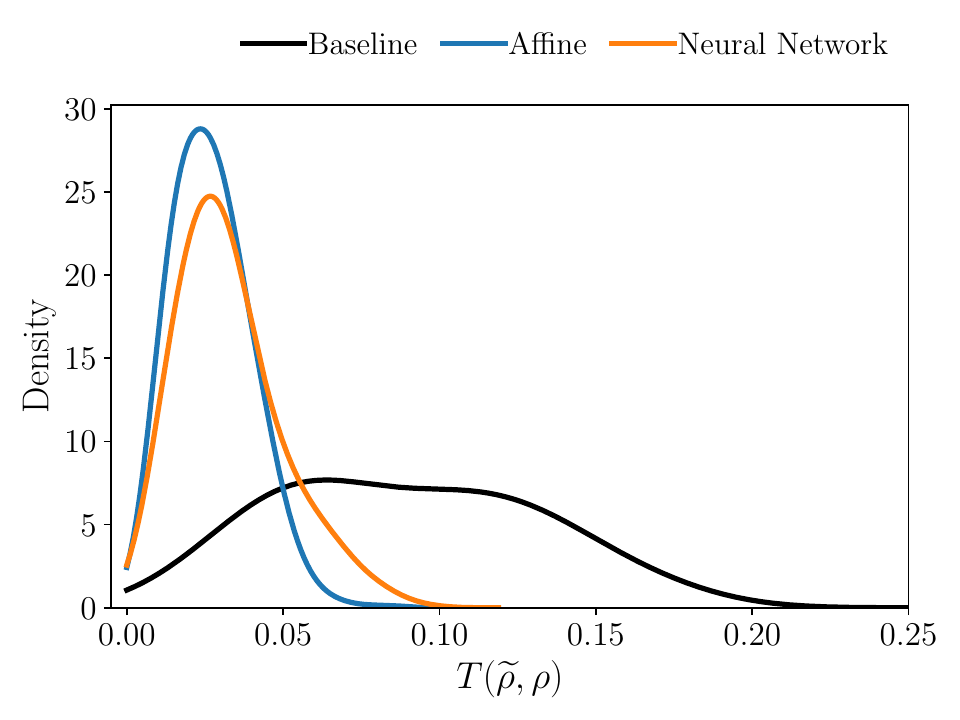} \caption{}\label{fig:NTCL:TrDist:Int}
\end{subfigure}
\begin{subfigure}[h]{0.33\textwidth} %
\includegraphicsifexists[width=\textwidth]{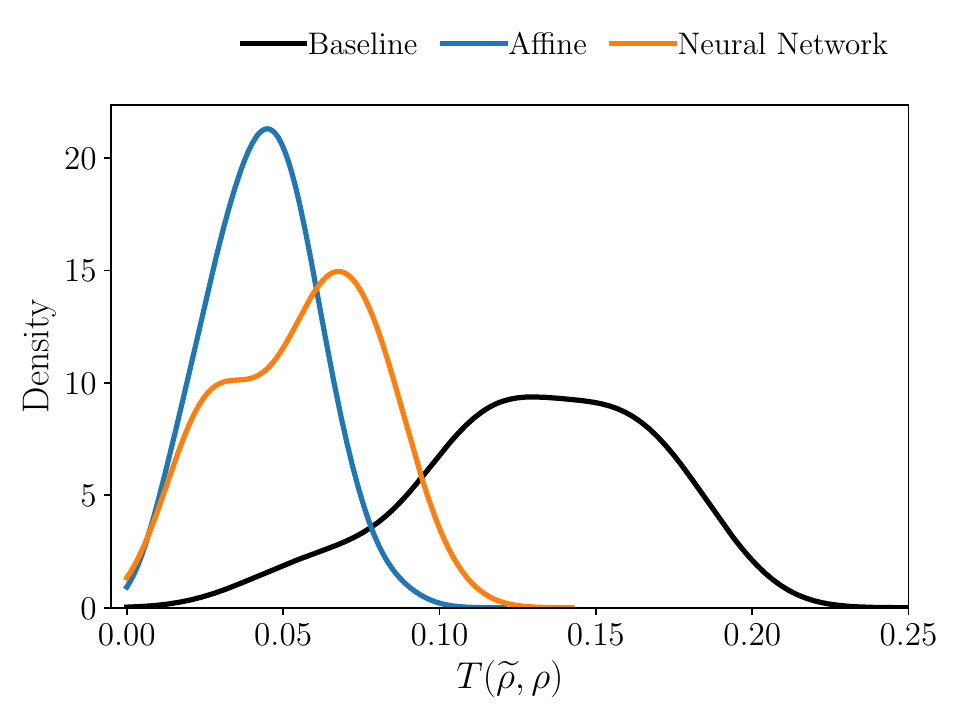} \caption{}\label{fig:NTCL:TrDist:Ext}
\end{subfigure}
\captionsetup{singlelinecheck=off,font=footnotesize}
\caption[]{Probability distribution of the trace-distance between the measured density matrices and those predicted by the (top) linear TCL and (bottom) nonlinear TCL master equations. (a,c) represent the trace distance from evolution within the training window and (b,d) represent the trace distance for evolution beyond the training window (i.e. extrapolation).}
\label{fig:TCL:TraceDistance}
\end{figure}

Figure \ref{fig:TCL:TraceDistance} shows the probability density of the trace distribution between the experimentally measured and predicted density matrices from evolution within (Figs. \ref{fig:TCL:TrDist:Int} and \ref{fig:NTCL:TrDist:Int}) and beyond (Figs. \ref{fig:TCL:TrDist:Ext} and \ref{fig:NTCL:TrDist:Ext}) the training window. We see that all models perform better than the master equation characterized via Ramsey interferometry. The higher densities for lower trace distances show that the KL-Sq. Exp. parameterization is more accurate for predicting dynamics within the training window but is outperformed by the affine model in extrapolating beyond the training domain. This can be attributed to the instability of the extrapolation due to higher order terms in the KL expansion. Furthermore, all models non-Markovian model outperform or are on-par with the Markovian Lindblad equation, with most significantly outperforming it. As before, the affine and neural network models for the linear and nonlinear TCL equations performed similarly in both interpolation and extrapolation. 

Table \ref{table:TCL:StatsMoment} shows the mean and standard deviation of the trace distance for different master eqautions and parameterizations. It shows that the affine model (for both TCL and nonlinear TCL) and the KL models yield lower means and standard deviation of the trace distances when compared to the Lindblad master equation, and significantly lower than the baseline model. The affine parameterization of the nonlinear TCL is on-par with the KL model for interpolation, but outperforms all other models in the remaining cases. 

\begin{table}[h]
\caption{Mean and standard deviation (in brackets) of the trace distance for different master equations and parameterizations in both interpolation and extrapolation.} \label{table:TCL:StatsMoment}
\centering
\begin{tabular}{c | l | c| c} 
 \hline
  &  & Interpolation & Extrapolation \\
 \hline 
 \hline 
  & Parameterization  & & \\
 \hline 
 \hline 
  Baseline & - & 0.09 (0.0423) & 0.13 (0.037)  \\ 
\hline 
\hline 
 Lindblad & - & 0.035 (0.016) & 0.063 (0.0254)  \\   
 \hline
 \hline 
 \multirow{4}{*}{Linear TCL} & \multicolumn{1}{l|}{Affine} & 0.028 (0.0128) & 0.047 (0.0178) \\
  & Neural Network & 0.038 (0.0172) & 0.056 (0.0242)  \\  
  & KL - Sq. Exp. & 0.025 (0.0122) & 0.052 (0.0214)  \\  
  & KL - Exp. & 0.031 (0.0146) & 0.058 (0.0233)  \\  
 \hline 
 \hline  
 \multirow{2}{*}{Nonlinear TCL} & \multicolumn{1}{l|}{Affine} & 0.027 (0.0126)  & 0.043 (0.0169) \\
  & Neural Network & 0.033 (0.0167) & 0.056 (0.0238)  \\  
 \hline 
 \hline
\end{tabular}
\end{table}

%% file: conclusion.tex
\section{Conclusion} \label{sec:Conclusion}

In this work, we developed a structure-preserving approach to characterizing non-Markovian quantum dynamics using the time-convolutionless (TCL) master equation, considering both linear and nonlinear formulations. To parameterize the master equation, we investigated two techniques: the Karhunen-Loève (KL) expansion and neural networks, leveraging experimental data from a superconducting qubit at the Quantum Device Integration Testbed (QuDIT) at LLNL. Our findings demonstrate that the KL expansion provided the most accurate predictions for the qubit’s dynamics within the training set. However, we observed that the instability of higher-order terms in the KL expansion led to a slight deterioration in extrapolation accuracy. This issue can be mitigated using L1 regularization, which promotes sparsity and improves generalization by reducing the influence of higher-order modes. Additionally, we found that the linear and nonlinear TCL master equations exhibited similar accuracy. Furthermore, our results indicate that both affine and nonlinear parameterizations of the master equation yielded similar accuracies, suggesting that the added complexity of nonlinear parameterization may not be necessary for single-qubit systems but could prove beneficial for modeling multi-qubit interactions and more complex noise processes. These insights provide a foundation for more efficient modeling strategies in open quantum systems and suggest promising directions for improving quantum control, error mitigation, and scalable system characterization in next-generation quantum devices